# Examination of evidence for collinear cluster tri-partition

Yu.V. Pyatkov[1,2], D. V. Kamanin[2], A. A. Alexandrov[2], I. A. Alexandrova[2], Z. I. Goryainova[2], V. Malaza[3], N. Mkaza[3], E. A. Kuznetsova[2], A. O. Strekalovsky[2], O. V. Strekalovsky[2], and V. E. Zhuchko[2]

[1]National Nuclear Research University MEPhI (Moscow Engineering Physics Institute), Moscow, Russia
[2]Joint Institute for Nuclear Research, Dubna, Russia
[3]University of Stellenbosch, Faculty of Military Science, Military Academy, Saldanha 7395, South Africa

**Background** In a series of experiments at different time-of-flight spectrometers of heavy ions we have observed manifestations of a new at least ternary decay channel of low excited heavy nuclei. Due to specific features of the effect, it was called collinear cluster tri-partition (CCT). The obtained experimental results have initiated a number of theoretical articles dedicated to different aspects of the CCT. Special attention was paid to kinematics constraints and stability of collinearity.

**Purpose**: To compare theoretical predictions with our experimental data, only partially published so far. To develop the model of one of the most populated CCT modes that gives rise to the so-called "Ni-bump".

**Method**: The fission events under analysis form regular two-dimensional linear structures in the mass correlation distributions of the fission fragments. The structures were revealed both at highly statistically reliable level but on the background substrate, and at the low statistics in almost noiseless distribution. The structures are bounded by the known magic fragments and were reproduced at different spectrometers. All this provides high reliability of our experimental findings. The model of the CCT proposed here is based on theoretical results, published recently, and the detailed analysis of all available experimental data.

**Results**: Under our model, the CCT mode giving rise to the Ni-bump occurs as a two-stage brake-up of the initial three body chain like the nuclear configuration with an elongated central cluster. After the first scission at the touching point with one of the side clusters, predominantly heavier one, the deformation energy of the central cluster allows the emission of up to four neutrons flying apart isotropically. The heavy side cluster and a dinuclear system, consisting of the light side cluster and the central one, relaxed to a less elongated shape, are accelerated in the mutual Coulomb field. The "tip" of the dinuclear system at the moment of its rupture faces the heavy fragment or the opposite direction due to a single turn of the system around its center of gravity.

**Conclusions**: Additional experimental information regarding the energies of the CCT partners and the proposed model of the process respond to criticisms concerning the kinematic constrains and the stability of collinearity in the CCT. The octupole deformed system formed after the first scission is oriented along the fission axis, and its rupture occurs predominantly after the full acceleration. Non-collinear true ternary fission and far asymmetric binary fission, observed earlier, appear to be the special cases of the decay of the prescission configuration leading to the CCT.

Detection of the $^{68\text{-}72}$Ni fission fragments with a kinetic energy $E < 25$ MeV at the mass-separator Lohengrin is proposed for an independent experimental verification of the CCT.

## I. INTRODUCTION

In recent publications [1–3], we have presented experimental indications of the possible existence of a new at least ternary decay channel of low excited heavy nuclei known as collinear cluster tri-partition (CCT). The bulk of the results has been obtained by using the "missing mass" approach. It means that two decay products (fragments) were detected in coincidence using a double armed time-of-flight spectrometer, while the significant difference between their total mass $M_s = M_1 + M_2$ and the mass of a mother system served as a sign of at least ternary decay. A fragment mass is calculated by the energy $E$ and the velocity $V$. Mainly a scattering of fragments at the entrance of an $E$–detector gives background events simulating ternary decay. Selection of the "true" events was provided by applying the gates on the fragments momenta, velocities, experimental neutron multiplicity, and the parameters sensitive to the fragment nuclear charge. Observation of the specific linear structures in the $M_1$–$M_2$ distributions (mass correlation plots) served as a criterion for a sufficient suppression of the background. Statistical reliability of the typical structures against a random background was estimated to exceed 98% [2]. The structures were reproduced at the spectrometers of two types. Earlier experiments were performed using gas filled detectors (modules of the FOBOS setup [4]). Later we switched to solid-state detectors, namely timing detectors, on the microchannel plates and the mosaics of PIN diodes (COMETA setup [2] and the similar ones [3]).

Even though mass reconstruction procedures for these two types of spectrometers strongly differ, the obtained results are in good agreement. All the structures revealed are somehow related to the magic fragments, such as $^{128}$Sn, $^{134}$Te, $^{72, 68}$Ni and the others. Thus, now we have an entire collection of different CCT manifestations, observed through the linear structures in the mass correlation distributions of the decay products [3].

Completely new results obtained in our experiments suggested independent experimental verification. To perform such an experiment at the mass-separator Lohengrin (ILL, Grenoble, France), estimation of the expected parameters of the CCT products was performed in the recent work [5]. The parameters should be compared with our experimental findings, only partially published so far.

## II. EXPERIMENTAL RESULTS

Below we discuss one of the most pronounced manifestations of the CCT called "Ni-bump", observed for $^{252}$Cf(sf) and in the reaction $^{235}$U($n_{th}$, f), using the detector modules of the FOBOS spectrometer (experiment Ex1) [4]. Fig. 1 shows the two-dimensional distribution ($M_2-M_1$) of the two registered masses of the coincident fragments in a logarithmic scale. The collinear fission events with a relative angle of $180\pm2^0$ fragments constitute the distribution. The "tails" in the mass distributions, marked 3−6 in Fig. 1(a), extend from the regions (1) and (2), which correspond to conventional binary fission. The tails are mainly caused by the scattering of the fragments on both foils, and/or on the grid edges of the "stop" avalanche counters, and/or in the ionization chambers. The existence of the small but important asymmetry should be emphasized in the experimental arrangement for two arms, which consists of the thin source backing (50 μg/cm$^2$ of Al$_2$O$_3$) of the target and the "start" detector foil, located only in arm1 (Fig. 1(a)). It is a marked difference in the counting rate, and in the shapes of the tails (3) and (4) that draws our attention. In the case shown in Fig. 1(a), there is a distinct "bump", marked (7), on top of the latter tail (4). The bump is located in the region corresponding to a large "missing" mass. We admitted the presence of the bump in arm 1 only due to the joint influence of the scattered medium and blocking grid on the initially collinear pair of the CCT products [1]. The statistical significance of the events in the structure (7) can be deduced from Fig. 1(b), in which the $M_1$ spectra (projection of the $M_2-M_1$ distribution onto $M_1$ axis) are presented in a comparative manner. For the experiment Ex2, performed at the modified FOBOS spectrometer [2], the difference spectrum between the tails (4) and (3) is depicted in the figure. The yield of the events in each difference spectrum is approximately $4 \times 10^{-3}$ relative to the total number of the events in the distribution shown in Fig. 1(a). A pronounced peak in the spectra in Fig. 1(b), centered on 68 u, is associated with a magic isotope of $^{68}$Ni, and that is why the bump (7) was called the "Ni- bump".

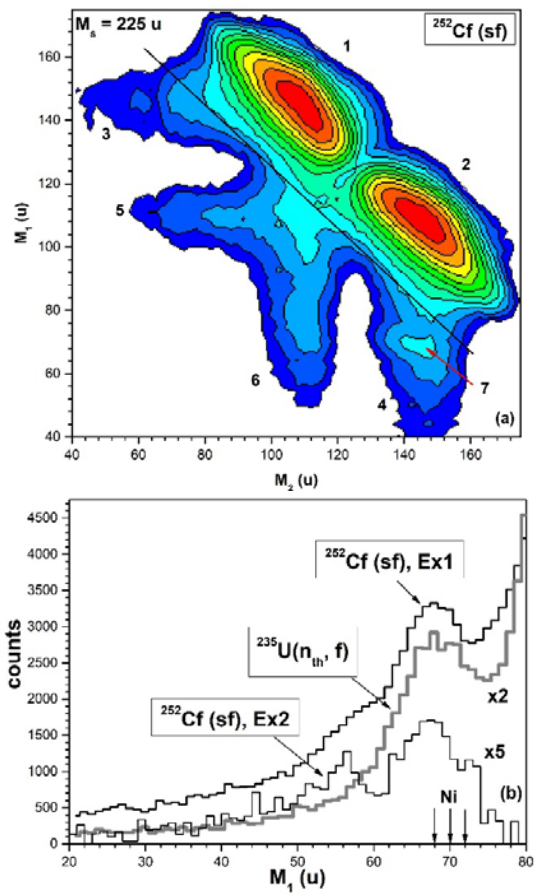

FIG. 1. (Color online). (a) Ex1: contour map (in logarithmic scale, the steps between the lines are approximately factor 2.5) of the mass–mass distribution of the collinear fragments of $^{252}$Cf (sf), detected in coincidence in the two opposite arms of the FOBOS spectrometer. The arrow marks a specific "bump" in arm1. (b) Projection of the "Ni-bump" onto $M_1$–axis obtained in three different experiments performed at the FOBOS spectrometer modules. It should be indicated that Fig. 1(a) was published in Ref. [1] whereas Fig. 1(b) in Ref. [2].

The internal structure of the Ni-bump was described in detail in the experiment Ex3 at the COMETA spectrometer [2]. This methodically quite different experiment confirmed our previous observations concerning the structures in the missing mass distributions. In this case, there is no tail caused by the scattering from the material in front of the $E$–detectors. Fig. 2(a) shows the region of the mass distribution for the fission fragments (FFs) from $^{252}$Cf (sf) around the Ni-bump ($M_1$ = 68–80 u, $M_2$ = 128−150 u). The structures are seen in the spectrometer arm facing the source backing only. No additional selection of the fission events was applied in this case, which resulted in the experiment having almost no background. A rectangular-like structure below the locus of binary fission is bound by magic nuclei (their masses are marked by the numbered arrows), namely $^{128}$Sn (1), $^{68}$Ni (2), and $^{72}$Ni (3). In Fig. 2(b), we demonstrate the projection of the linear structure seen at the masses of 68 and 72 u. Two tilted diagonal lines with $Ms$ = 196 u and $Ms$ = 202 u (marked by number 4) start from the partitions 68/128 and 68/134 (all the nuclei are magic) respectively. The discussion of these structures is beyond the scope of this paper.

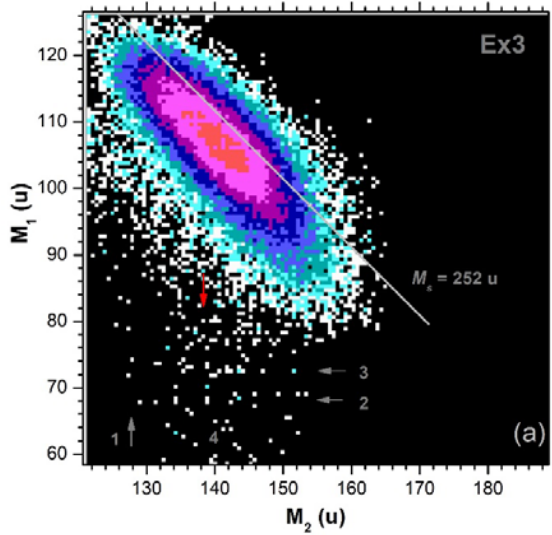 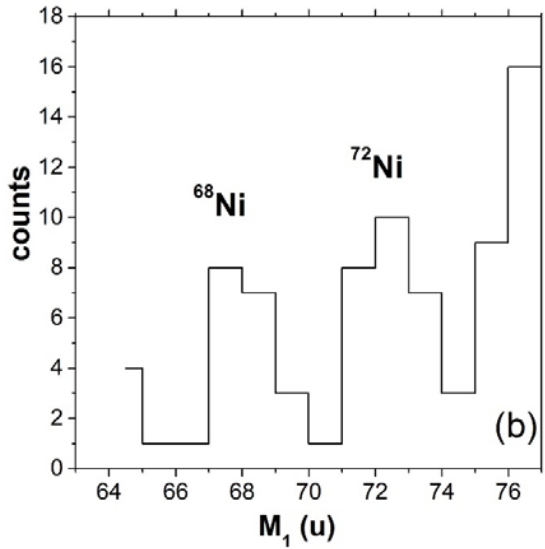

FIG. 2. (Color online). Ex3: region of the mass–mass distribution for the FFs from $^{252}$Cf (sf) around the Ni-bump (similar to that marked by an arrow in Fig. 1(a)). It should be indicated that Fig. 2(b) was published in Ref. [2].

In fact, only two fragments were detected in each decay event. The mass and velocity of the "missed" fragment were calculated based on the laws of mass and momentum conservation. In each event showing the missing mass (ternary event), we mark the masses of the fragments in order of their decreasing masses $M_H$, $M_L$ and $M_T$ (Ternary particle) respectively. Fig. 3(a) demonstrates a correlation between the velocities of two lighter partners of the ternary decay. Only the events for which $M_L$ = (67–75) u (Ni-peaks in Fig. 2(b)) are under analysis. Their total yield does not exceed 2.5x10$^{-4}$ per binary fission. Three different groups of events are vividly seen in the figure. They are marked by the signs $w1$–$w3$ respectively. The panels illustrating the decay scenario will be discussed below. The energy spectrum of the detected Ni nuclei is depicted in Fig. 3(b). The energy correlations $E_L$–$E_T$ and $E_L$–$E_H$ are shown in Figs. 3(c) and 3(d).

Even though, there are only a few points on the line 1 ($M_1 \approx$ 128 u) in Fig. 2(a), the energy spectrum of the Sn fragment clearly shows two peaks, centered respectively at $E_2 \sim$ 20 MeV (predominantly) and $E_2 \sim$ 90 MeV.

The spectrum of the excitation energy in the scission point $E_{ex}$ = $Q_3$–TKE$_3$ (where $Q_3$ is an energy released in ternary decay, TKE$_3$ is a total kinetic energy of three decay partners) is presented in Fig. 4.

The region of the Ni-bump, almost free from the background, was observed in Ex2 (Fig. 5(a)) due to the application of the gate on the experimental neutron multiplicity $n$ = 2 and an additional gate in the $V_1$–$E_1$ distribution [2]. Real multiplicity was estimated to be approximately four for isotropic neutron source. The lines corresponding to the magic isotopes of $^{68, 72}$Ni (marked by arrows 1 and 2) are seen. From the left side, the structure is bounded by a magic $^{128}$Sn fragment (its mass is marked by arrow 3 on the $M_2$ axis). The energy spectrum of these fragments is shown in Fig. 5(b). Good "cleaning" of the distribution from the background in Fig. 5(a) allows observing the lightest partners of the ternary decay. The tilted lines at the bottom of Fig. 5(a) marked by numbers 4–8 correspond to the magic "missing" fragments with mass numbers: 68, 72, 80, 85, and 102 respectively.

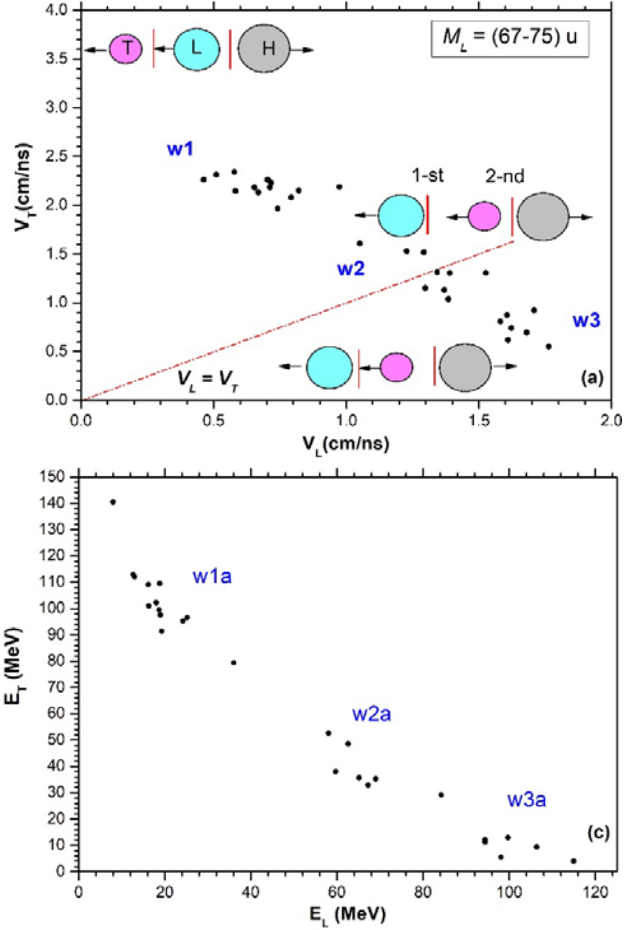
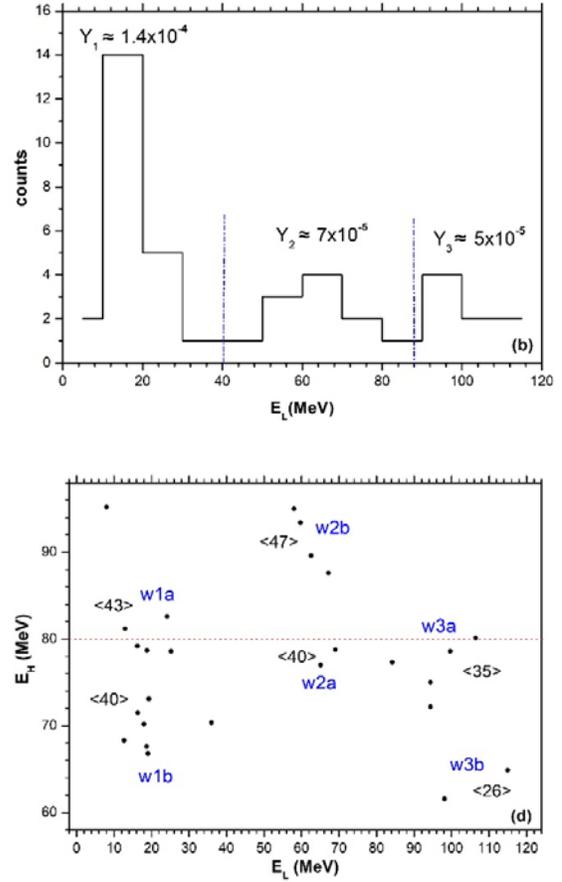

FIG. 3. (Color online). Ex3: velocities $V$ and energies $E$ for the ternary events with $M_L$ = (67–75) u (Ni-peaks in Fig. 2(b). Correlation between the velocities of two lighter partners of the ternary decay – (a), energy spectrum of the detected Ni nuclei (the yields per binary fission are marked above each peak) – (b), energy correlations $E_L$–$E_T$ and $E_L$–$E_H$ – (c) and (d) respectively. The sketches in the panels illustrate the decay scenario to be discussed below. See the text for details.

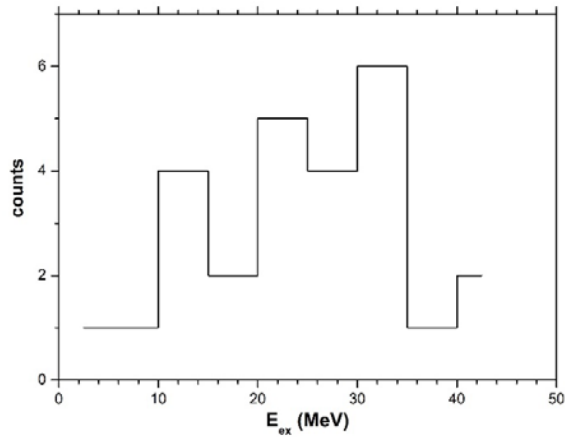

FIG. 4. Ex3: the spectrum of excitation energy in the scission point $E_{ex} = Q_3 - TKE_3$ for the events presented in Fig. 3.

For the events on lines 4 and 5, $E_T$ = 30–50 MeV, that is compatible with the energy of ternary particles from the group $w2a$ in Fig. 3(c). The energy spectrum of the Ni fragments (Fig. 5(b)) is also compatible with the most energetic peak in Fig. 3(b). Here, the absence of two other peaks with the energies of approximately 60 and 20 MeV can be explained correspondingly by the selection with the gate (see Fig. 9(a) in Ref. [2]) in the $V_1$–$E_1$ distribution, and missing of the low energy Ni ions in the foils of the FOBOS modules used in Ex2.

Experimental information regarding the energies of the CCT partners presented in this section allowed us to formulate an adequate scenario of tri-partition and chose appropriate parameters of a prescission configuration (Sec. IV).

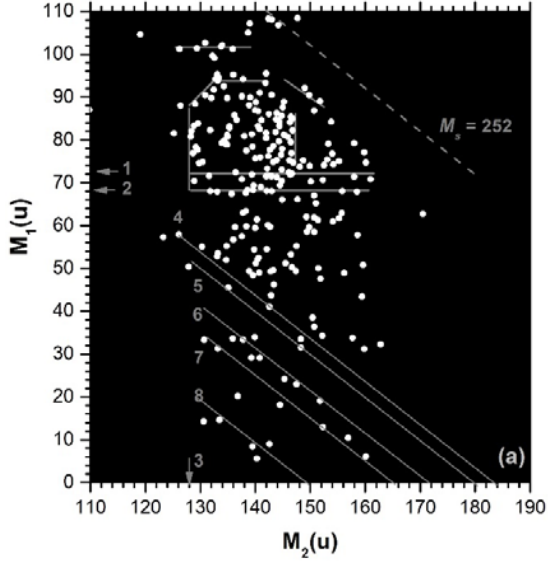
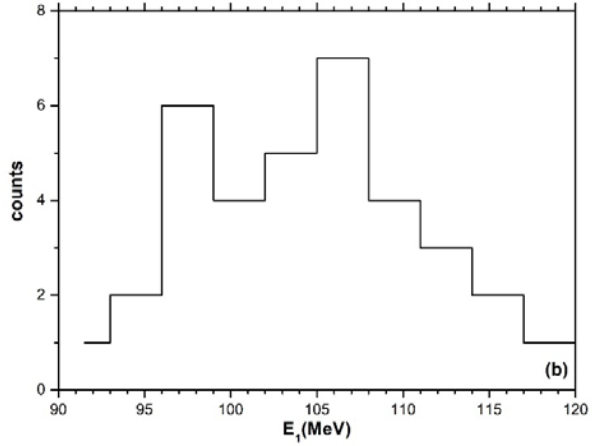

FIG. 5. Ex2: (a) results for experimental neutron multiplicity, $n = 2$, the mass-mass distribution of the FFs, and an additional gate in the $V_1$–$E_1$ plot [2]. The lines are drawn to guide the eye. (b) Energy $E_1$ for $^{68,72}$Ni nuclei (marked by arrows 1 and 2 in Fig. 5(a)). It should be indicated that Fig. 5(a) was published in Ref. [2]. See the text for more details.

## III. THEORETICAL BACKGROUND

Among all the theoretical articles initiated by the results of the experiments mentioned above, article [6] deserves special attention. The theoretical model of Ref. [6] was described in detail in Refs. [7, 8]. Under the three-center shell model the potential energy surfaces for few ternary combinations in a fission channel were calculated for the $^{252}$Cf nucleus. The fission barrier for the $^{132}$Sn+$^{48}$Ca+$^{72}$Ni ternary splitting is shown in Fig. 6. According to the figure, the exit point corresponds to R ~ 22.4 fm, i.e. elongation of the system exceeds the length of the configuration of three touching spheroids. If just a Ca nucleus took upon itself all extra elongation, the axis ratio of the corresponding spheroid would be approximately 1:1.6. In fact, as it is the case with heavy actinides, a scission point can be reached after descent from the barrier at larger elongations. Another point to be stressed is that the maximal total shell correction of three magic nuclei, forming the prescission chain (~10 MeV), exceeds the corresponding value, even for the double magic $^{208}$Pb nucleus (~9 MeV [9]). Thus, the elongated shape of the nuclear system minimizing the Coulomb energy and the big shell correction give rise to the valleys of true ternary fission, revealed in the paper under discussion.

In the work of Ref. [10], an approach of the trinuclear system (TNS) was applied for the analysis of the spontaneous ternary fission of $^{252}$Cf. Both the Coulomb and nuclear forces between decay partners were taken into account. *The stage that precedes the formation of the TNS is not studied* [11]. Selected results of Refs. [10, 11] are displayed in Fig. 7.

The scheme showing the variables used in the calculation is presented in the inset in Fig. 7(a). The potential energy $V(R_{12}, x3, y3 = 0)$ of the TNS as a function of $x3$ at different values of $R_{12}$ is shown in Figs. 7(a) and 7(b). The potential energy surface $V(R_{12}, x3, y3)$ at the value of $R_{12} = (22, 24)$ fm is presented in Figs. 7(c) and 7(d).

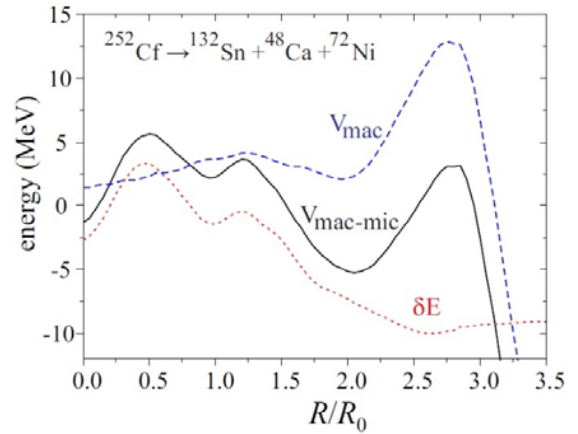

FIG. 6. (Color online). Macroscopic potential energy (dashed line), shell correction (dotted line), and total macro-microscopic potential energy (solid line) of the $^{252}$Cf nucleus corresponding to the $^{132}$Sn + $^{48}$Ca + $^{72}$Ni ternary splitting [6]. Here $R$ is an approximate distance between the mass centers of the side fragments, $R_0 = 1.16A^{1/3}$ is a radius of a mother nucleus.

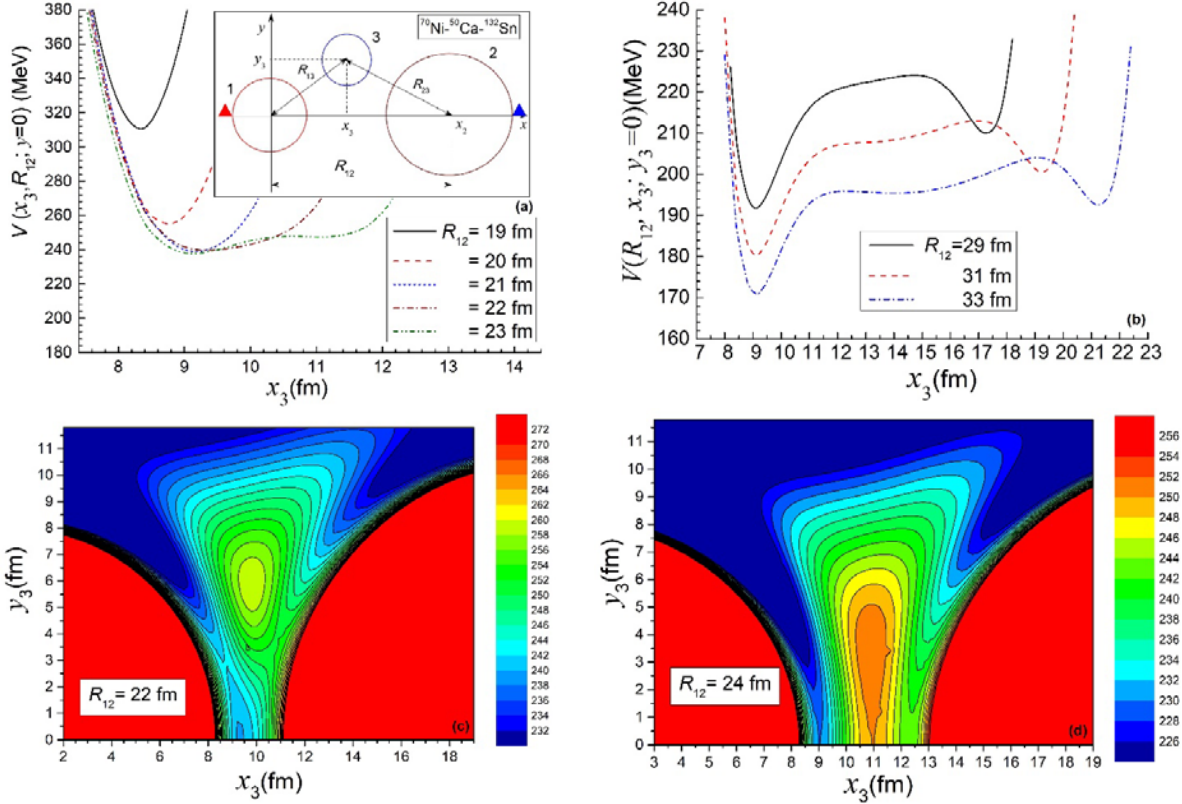

FIG. 7. (Color online). The potential energy $V(R_{12}, x3, y3 = 0)$ of the TNS as a function of $x3$ at different values of $R_{12} = 19, 20, 21, 22, 23$ fm. In the inset – the scheme showing the variables used in the calculation [10] of the potential energy $V(R_{12}, x3, y3)$ (a). The same is true for the $R_{12} = 26, 27, 28$ fm (b). The potential energy surface $V(R_{12}, x3, y3)$ as a function of the position $x3$ and $y3$ of the center-of-mass of the middle fragment "3" (Ca) at the value of $R_{12} = 21$ fm (the relative distance between centers of fragments' masses "1" and "2") (c) and of $R_{12} = 24$ fm (d).

At the distances $R_{12} < 24$ fm, there is a potential pocket between the side nuclei $^{70}$Ni and $^{132}$Sn for the third fragment of $^{50}$Ca (see, for instance Fig. 7(c)). According to the authors of Ref. [10], with further elongation starting from $R_{12} = 24$ fm (Fig. 7(d)), "there is no minimum for Ca in the valley around Ni at $x3 > 0$ fm, $y3 > 0$ fm". This conclusion is only formally true if the lightest fragment ($^{50}$Ca) is supposed to be located exclusively *between* the side heavy fragments (see the inset in Fig. 7(a)). The analysis in Refs. [12, 13] shows also that "the collinear geometry with *the lightest fragment at the center* between two heavier nuclei is expected to give the highest probabilities in the decay". This conclusion is definitely valid only for a *prescission configuration* of the decaying system. The situation changes after the first rupture. Indeed, there is a potential *valley along the surface* of a Ni nucleus with the bottom sloped to the potential minimum in the point that we marked by a triangle in red in the scheme in Fig. 7(a). The slope of the valley bottom is evidently due to the decrease of the Coulomb energy contribution from Sn with increase of the distance $R_{32}$. The similar potential minimum on the surface of Sn is marked by a triangle in blue.

The possible fission channels into fragments of similar size are predicted from potential-energy surface (PES) calculations in Ref. [14]. These PES's show pronounced minima for several ternary fragmentations involving magic nuclei. Statistical fission process is considered; and the observed yield of the distinct CCT mode is supposed to be proportional to its *phase space*. Rectangular structures in the experimental mass-mass distributions of the FFs (Figs. 3 and 5 in Ref. [14]) are treated on a qualitative level as the direct manifestation of the PES-landscape. In fact, the symmetry of the experimental structures observed is only due to the symmetry of the spectrometer arms. The experimental data discussed in Ref. [14] are still waiting for an adequate theoretical interpretation.

An original approach to description of the CCT was proposed in Ref. [15]. It is quite different from TNS concept and generically involves fission into three fragments. The mechanism in question is driven by *a hexadecapole deformation of the fissioning nucleus* and its scission could occur by means of tunneling of two identical side clusters through the very high Coulomb barriers. How realistic this approach is depends on the probability of such fission way to be estimated at least in order of magnitude.

The detailed analysis of the CCT regarding kinematic constraints and stability of collinearity was proposed in Ref. [5]. Three different models were under consideration. The "sequential" decay model is based on two sequential binary fissions, with long timescales between successive scissions. The "almost sequential" decay model suggests the formation of partially accelerated fragments before the second scission. No intermediate steps or fragments are expected under the "true ternary" decay model.

The FFs energies found under the sequential decay model (Fig. 9(b) in Ref. [5]) are identical to the ones calculated by Vijiayaraghavan et al. [16] (see Fig. 6 of Ref. [16]). In both works, a very artificial way was used to introduce the excitation energy of the intermediate fragment $E^*_{IF}$, forming after the first rupture. This energy is subtracted from the energy going into the fragments' kinetic energies at the first step (formula (8) in Ref. [16]), but then it is added to the kinetic energies of the fragments forming at the second step (formula (16) in Ref. [16]). Such approach does not violate the law of energy conservation, *but the full conversion of the excitation energy of the fissioning system into the kinetic energy of fragments contradicts the well-known experimental facts* [17]. As a result, under the model, regardless of the $E^*_{IF}$ value, the predicted total kinetic energy of all three fragments, TKE$_3$, stays constant and equal to the reaction energy. Even if one ignores the incorrect way of introducing $E^*$, the resultant "true cold ternary fission" is extremely improbable or forbidden. An increase of the energies of Ni and Ca with increase of $E^*_{IF}$ (Fig. 8(a)) could be regarded as a model artefact. Based on Fig. 8(a), the conclusion of the authors of Ref. [5]: "If events are found above the maximum energy of binary fission (shown by the red lines), the origin must be ternary fission." is warped and cannot be a criterion for searching for the ternary fission events. It should be stressed that the criticism above is the only way of taking into account $E^*_{IF}$, but it is not the idea of sequential two-step ternary decay.

Under the "true ternary" decay model, the total excitation energy TXE of a decaying system is subtracted from $Q$-value, and the difference defines the kinetic energies of the fragments according to the laws of energy and momentum conservation. The total kinetic energy decreases with an increase of TXE (Fig. 8(b)).

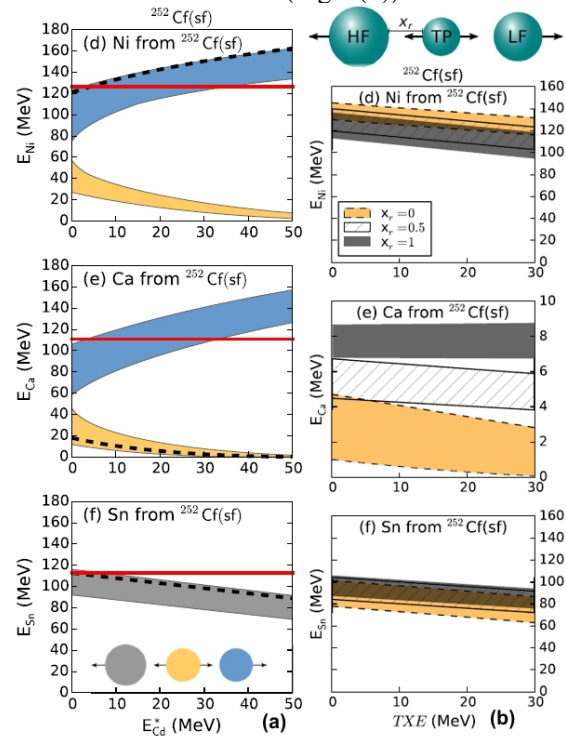

FIG. 8. (Color online). Areas of attainable final fragment kinetic energies versus the excitation energy of the intermediate fragment in the sequential decay $^{252}$Cf (sf) $\to$ Sn + Cd $\to$ Sn + Ca + Ni. The colors indicate formation position, as shown by the inset (a). Similar graphs but for the true ternary decay $^{252}$Cf (sf) $\to$ Sn + Ca + Ni. Each figure is a specific element, and different areas indicate the choice of $x_r$, as shown in the upper part of the figure (b). Contact position of the TP with LF corresponds to $x_r$ = 1 (appropriate regions are shown in black) [5].

At the end of this section, there should be mentioned a group of the theoretical articles dedicated to the prediction of the absolute yields of some CCT modes. We cover them in chronological order.

Under the three-cluster model within a spherical approximation and satisfying the condition on the FF masses $A1 \geq A2 \geq A3$ independent and overall relative yields are calculated for the partitions with $A3$ = 1 to 84 [18]. The obtained estimations agree poorly with the known experimental data. For instance, the ratio of the yields $Y(^4\text{He})/Y(^{10}\text{Be})$ differs from the experimental one Ref. [19] by at least six orders of magnitude (Fig. 5 in Ref. [19]). In the next work of the same authors, this difference is somewhat smaller (Fig. 10 in Ref. [20]) but still too big to unconditionally believe the conclusions of the article.

In the work [21], the yield of ternary fission products is calculated using the statistical model based on the driving potentials for the fissionable system. The asymmetric fission channel as the first stage of a sequential mechanism is supposed. The conclusion about the agreement of the theoretical results of the yield of $^{80}$Ge and $^{84}$Se isotopes with the experimental data is likely based on a misunderstanding of these data. Predicted yield for the partition $^{82}$Ge-$^{72}$Ni-$^{82}$Ge (Table II in Ref. [21]) is approximately $3\times10^{-4}$, while the corresponding experimental value does not exceed $1.6\times10^{-6}$ (Fig. 7(d) in Ref. [22]).

"Almost sequential" mechanism of true ternary fission is analyzed in detail in Ref [23]. The authors consider the collinear sequential ternary decay with a very short time between the ruptures of two necks connecting the middle cluster of the ternary nuclear system and outer fragments. The quadrupole deformation parameters of the first-excited 2+ state of nuclei are used in calculation of the PES. A probability of approximately $10^{-3}$ per binary fission is obtained for the yield of clusters such as $^{70}$Ni, $^{80-82}$Ge, $^{86}$Se, and $^{94}$Kr in the ternary fission of $^{252}$Cf. The yield agrees with the experimental one at least for the Ni clusters (Fig. 3(b)), taking into account some part of the experimental yield below the experimental energy threshold. A crucial test for the adequacy of the adopted model would be the evaluation of both the yields and energies of the fragments in the unified approach for the comparison with the experimental data (Fig. 3).

Three collinear touching deformed fission fragments are supposed to constitute a configuration of the system after scission of necks in a three-fragment fission [24]. Initial nucleus undergoes ternary fission through the lowest barrier; and the yield of the three-fragment partition is proportional to the total fragment intrinsic energy $E^*$ at the lowest barrier point. The calculated values of the absolute yields for the light particles Be÷Si agree with the experimental data. At the same time, the partition into three magic clusters $^{72}$Ni+$^{48}$Ca+$^{132}$Sn, rated as the most favorable for $^{252}$Cf (sf) in Refs. [6, 10], is forbidden in the approach under discussion because $E^* < 0$. The authors note that similar fission channels "may appear in the framework of other mechanisms of the three-fragment formation".

Thus, in most reviewed theoretical works dedicated to true ternary fission and the CCT the concept of a dinuclear or trinuclear system (TSM) is implemented. The stage that precedes the formation of a prescission configuration is not studied [11], what provides a principal uncertainty in initial conditions for the further analysis. In contrast, the calculations based on the three-center shell model allow tracing of the shape of the decaying nucleus in the valley of true ternary fission from the ground state up to the exit point from under the fission barrier [6]. We combine both approaches in our scission point calculations (Sec. IV), aimed at reproduction of experimental FF energies (Fig. 3).

## IV. SCISSION POINT CALCULATIONS

The following scenario of the CCT process can be proposed based on our experimental findings and recent theoretical calculations. According to Ref. [6], the exit point from under the barrier in the potential valley leading to the $^{132}$Sn + $^{48}$Ca + $^{72}$Ni ternary splitting (Fig. 6) corresponds to a much more elongated configuration in comparison with the chain of three touching spherical nuclei. The distance $R_{12}$ between the centers of the side clusters was estimated to be above ~23 fm. Likely, the central fragment (Ca) takes upon itself almost all extra elongation. After a rupture occurs, for instance on the boundary of Ca and Sn clusters, the Ni cluster very quickly (in comparison to full acceleration time) attracts the Ca "neck". Part of the released deformation energy is spent on emission of neutrons flying apart isotropically. Thus, the formed pear-shaped Ni-Ca dinuclear system can rotate around the center of its gravity by $180^0$, so that the Ca "tip" appears in the position marked by a red triangle in Fig. 7(a). Such orientation is the most energetically favorable. Octupole vibrations [25] could be another reason for the change in the orientation of the "tip". Formation of the Ca-Sn system with similar features is less probable [10], but if it forms after the first rupture, the Ca tip can turn to the position marked by a blue triangle in Fig. 7(a).

The formed dinuclear system can evolve towards fusion or rupture. In the first case, we deal with a binary fission of a mother nucleus, and in the second instance with a ternary fission.

The key propositions of the CCT scenario above could be detailed to explain the experimental results presented in Fig. 3. It is noteworthy that each group of events $w1$-$w3$, seen in Fig. 3(a), consists of two subgroups in plot $E_H$–$E_L$ (Fig. 3(d)), which differ by the mean mass of the lightest cluster (shown in brackets) and the energies of the side clusters.

Presumable decay scenarios for all subgroups are presented in Fig. 9. A precission configuration of the system is demonstrated in the third column of the figure. For all the cases fission fragment FF$_1$ is supposed to be $^{70}$Ni, the mass of the FF$_2$ corresponds to the mean mass of the lightest cluster (shown in brackets in Fig. 3(d)), and the mass of the heavy cluster is calculated using the law of mass conservation. The FFs charges are calculated according to the hypothesis of unchanged charge density. Configurations of the system after the first and the second ruptures are shown respectively in the fourth and the fifth columns of the table. Parameters of the calculations and the results obtained are presented in Table I.

The following text comments on each row of the Fig. 9.

Row № 1. After the first rupture, the formed pear-shaped dinuclear system rotates around the center of its gravity by $180^0$, acquiring more energetically favorable position. It can also happen during the acceleration due to octupole vibration of the dinuclear system. A low energy of the $FF_2$ is due to its deceleration after the second rupture.

Row № 2. The fragments $FF_1$ and $FF_3$ in the events from the group $w2$ (Fig. 3(a)) have very close velocities, which cannot occur if the second rupture takes place in the dinuclear system $FF_1$–$FF_3$. In contrast, the forming of the dinuclear system $FF_3$–$FF_2$ after the first rupture should be considered.

Row № 3. Small velocities of the $FFs_2$ in the locus $w3$ (Fig. 3(a)) are assumed to be due to deceleration after the second rupture. This sets the order of the fragments $FF_1$ and $FF_2$ at the scission.

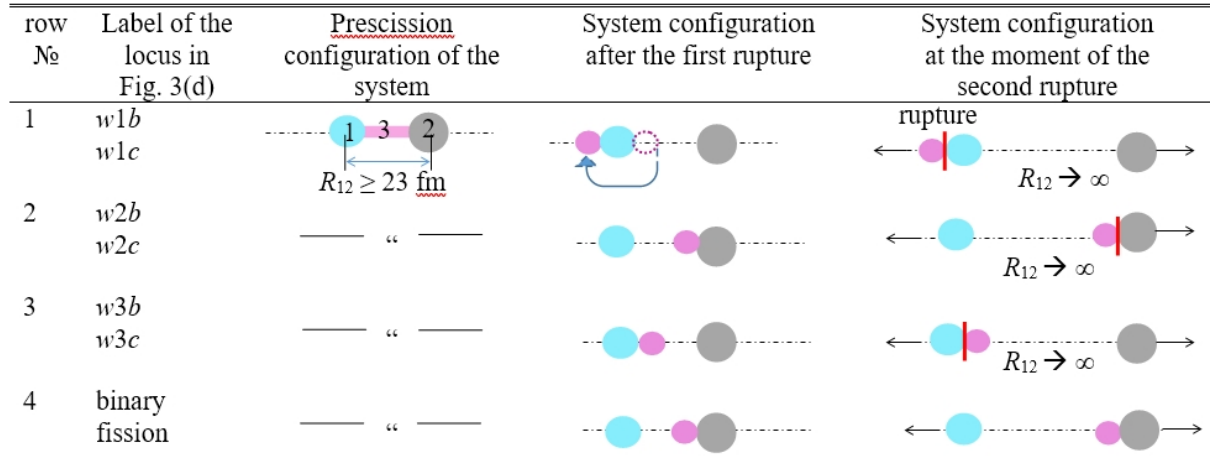

FIG. 9. (Color online). Pictograms illustrating scenarios of different CCT modes observed in the experiment. See the text for details.

TABLE I. (Color online). Results of the model calculations. Ternary partitions close to the experimental ones and based on magic constituents (marked in bold) are shown in square brackets. See the text for details.

| № | locus | nucl. configuration | $R_{12}$, fm | $E_{Hsec}$, MeV | $E_H$, MeV | $V_L$, cm/ns | $V_{Lsec,}$ cm/ns | $V_T$, cm/ns |
|---|---|---|---|---|---|---|---|---|
| 1 | $w1b$ | $^{70}$Ni – $^{43}$S – $^{139}$Xe [**$^{70}$Ni** – **$^{42}$S** – **$^{140}$Xe**] | ≤ 27 | 71 | 80.4±1.8 | 0.71±0.1 | | 2.16±0.06 |
| 2 | $w1c$ | $^{70}$Ni – $^{39}$Si – $^{143}$Ba [**$^{70}$Ni** - **$^{38}$Si** – **$^{144}$Ba**] | ≤ 30 | 58 | 69.5±2.6 | 0.68±0.06 | | 2.19±0.13 |
| 3 | $w2b$ | $^{70}$Ni – $^{47}$Ar – $^{135}$Te [**$^{72}$Ni** – **$^{46}$Ar** – **$^{134}$Te**] | ≤35 | | 91.4±3.1 | 1.30±0.06 | 1.34 | 1.33±0.22 |
| 4 | $w2c$ | $^{70}$Ni – $^{40}$S – $^{142}$Xe [**$^{70}$Ni** – **$^{42}$S** – **$^{140}$Xe**] | ≤35 | | 77.9±1.3 | 1.36±0.03 | 1.34 | 1.31±0.004 |
| 5 | $w3b$ | $^{70}$Ni – $^{35}$Al – $^{147}$La [**$^{70}$Ni** – **$^{34}$Mg** – **$^{148}$Ce**] | ≤ 32 | 60 | 76.6±3.1 | 1.62±0.04 | | 0.78±0.08 |
| 6 | $w3c$ | $^{70}$Ni – $^{26}$Ne – $^{156}$Nd [**$^{70}$Ni** – **$^{28}$Ne** – **$^{154}$Nd**] | ≤ 28 | 52 | 63.2±2.3 | 1.68±0.1 | | 0.58±0.05 |
| 7 | bin. fiss. | $^{70}$Ni – $^{50}$Ca / $^{132}$Sn $^{182}$Yb | | | TKE 141 MeV $E_{Ni}$ = 102 MeV | | | |

Row № 4. A significant feature of the prescission ternary configuration used above is that for the binary fission, predicted via this configuration, TKE ~ 140 MeV, which agrees well with the experimental value calculated in

Refs. [26, 27] for far asymmetric binary fission, with Ni as a light fragment.

The results of the calculations are compared with the experimental ones in Table I. Mean values of $E_H$, $V_L$, $V_T$ with the standard deviations from them were calculated using experimental data for each locus $w1b \div w3c$ in Fig. 3(d). Strictly speaking, it is not evident that these mean values would be self-consistent, namely satisfy the laws of energy and momentum conservation. And so, the following approach to testing has been developed.

In case of ternary decay, after full acceleration (at infinity) of all the fragments, both laws of energy and momentum conservation should be met:

$$E_1 + E_2 + E_3 = Eint \quad (1)$$
$$\vec{p_1} + \vec{p_2} + \vec{p_3} = 0$$

where $Eint$ is the interaction energy between the fragments at the beginning of the acceleration; $E_i$ and $p_i$ are respectively their energies and momenta. Thus, there are three unknown velocities, and one has only two equations for their determination. However, changing step-by-step one of the velocities or energies, for instance $E_H$, we can solve the set of Eq. (1) for each fixed value of $E_H$. The result of such calculations for the locus $w3b$ (see Table I and Fig. 9) is shown in Fig. 10. According to the algorithm, any vertical line intersecting both the $E_H$ axis and the curves above provides a trio of parameters, namely $\{E_H, V_L, V_T\}$, satisfying the system (1).

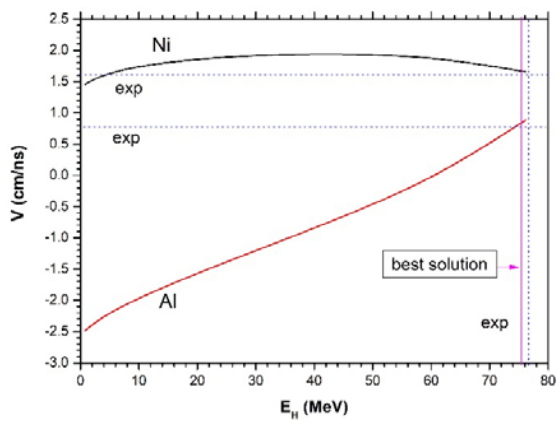

FIG. 10. (Color online). Velocities $V_L$, $V_T$ and energy $E_H$ of the ternary fragments satisfying the system (1). Calculations were performed for the locus $w3b$. See text for details.

Under this approach the mean values of $E_H$, $V_L$, $V_T$ are reproduced with approximately a five per cent margin of error at $Eint = \text{TKE}_{exp}$. Estimated $R_{12}$ values based on $\text{TKE}_{exp}$ for each configuration of the system after the first rupture (Fig. 9) are also shown in Table I. We varied $R_{12}$ in order to satisfy the equality

$$E_b + E_C = \text{TKE}_{exp},$$

where $E_b$ is the interaction energy of the constituents of the dinuclear system, formed after first rupture at the Coulomb barrier; $E_C$ is the energy of the Coulomb interaction of these constituents with the side fragment. Knowing $R_{12}$ one can estimate the energy $E_{Hsec}$ (or velocity $V_{Lsec}$) of the fragment separated first under condition of sequential fission. By definition, within this model, the second rupture occurs after the full acceleration at $R_{12} \to \infty$. The estimated values are approximately 80% of the experimental ones. This indicates that the real ternary fission should be treated as 'almost sequential" but it is very close to the sequential one. It should be noted that the rotation of the dinuclear system during acceleration (Row №1 in Fig. 9) greatly reduces actual $E_{int}$. We mark this condition with the sign ($\leq$) for the corresponding $R_{12}$ values in Table I.

## V. RESPONSES TO SOME CRITICAL QUESTIONS

There are some obvious questions concerning the nature of the CCT process and specificity of its observation in an experiment. We propose our current understanding of both.

### A. Why such a high total yield of the CCT is observed in comparison with conventional ternary fission?

#### A.1. Conventional ternary fission.

Ternary decay of the heavy nucleus with emission of predominantly α-particles, in the plane approximately perpendicular to the fission axis, is known as conventional ternary fission or just ternary fission. Such decay was observed for the first time in 1947 in the photo emulsion, containing a uranium salt, after irradiation by slow neutrons [28].

The characteristics of the angular distributions point to emission of α-particles from a region between the two nascent fragments during the neck rupture. The yield ratio of ternary/binary fission (t/b) in thermal neutron-induced fission stays close to $2 \times 10^{-3}$ for reactions, ranging from $^{229}\text{Th}(n_{th}, f)$ up to $^{251}\text{Cf}(n_{th}, f)$. In spontaneous fission, the t/b ratios are slightly larger [29]. For the reaction $^{249}\text{Cf}(n_{th}, f)$, the heaviest isotopes detected at the mass-separator Lohengrin were $^{37}\text{Si}$ and $^{37}\text{S}$ (t/b ≈ $10^{-9}$) [30].

Remarkable conclusions concerning the mechanism of ternary fission were put forward in the theoretical works [31, 32]. The relative yields of various light charge particles (LCP) are determined by the formation probability of the LCP and the likelihood of ternary decay. Formation probability is proportional to the spectroscopic factor. By definition, the spectroscopic factor is the weight of a certain binary configuration in the wave function of the nucleus. Table II demonstrates a revealed correlation between the formation probabilities $S$ and experimental yields $Y_{exp}$ of different LCPs.

Thus, one can conclude that the probability of ternary decay depends weakly on the type of the LCP. And hence, the yields of different LCPs are ruled by their formation probabilities $S$. The latter was estimated as $S(^{14}C) = S(_{4He})^3$, $S(^{20}O) = S(_{4He})^4$, and so on. Based on the arguments above, the authors came to a conclusion that "the sequential formation of the LCP from the correlated $^4$He in the region between two heavy fragments looks realistically". Analyzing the possibility of the multicluster accompanied fission of $^{252}$Cf, the authors of Refs. [33, 34] also pointed out that "the most favorable mechanism of such a decay mode should be the emission from an elongated neck formed between the two heavy fragments". Emission of light clusters accompanied fission, e.g., α–particles, $^{10}$Be, $^{14}$C, $^{20}$O, or combinations of them were studied.

TABLE II. Correlation of the formation probabilities $S$ and experimental relative yields $Y_{exp}$ of different LCPs. The values of $S$ and $Y_{exp}$ are given with respect to those for $^4$He [31].

| LCP | $Y_{exp}/Y_{exp}(^4He)$ | $S/S_{4He}$ |
|---|---|---|
| $^4$He | 1 | 1 |
| $^7$Li | 5x10$^{-3}$ | 4.2x10$^{-3}$ |
| $^{10}$Be | 1.3x10$^{-2}$ | 5x10$^{-2}$ |
| $^{11}$Be | 6x10$^{-4}$ | 2.1x10$^{-4}$ |
| $^{14}$C | 5x10$^{-3}$ | 2.5x10$^{-3}$ |
| $^{20}$O |  | 1.3x10$^{-4}$ |

Following this approach, a rapid decrease of the LCP yield with an increase of their masses is explained by the power-law decrease of the formation probabilities $S$.

### A.2. CCT mechanism.

As was noted in Sec. V, the same elongated prescission configuration of a decaying nucleus can lead both to the CCT and to far asymmetric binary fission with the TKE ~ 140 MeV. Low values of the TKE are directly correlated with increased multiplicity ν of the fission neutrons [35], confirming the concept of the cold deformed fission [36]. According to the experimental data [37, 38], an emission of even eight neutrons for $^{252}$Cf (sf) exceeds 10$^{-3}$ (Table III), i.e. *highly deformed prescission configurations occur with the probability comparable to the total yield of the CCT events.*

Calculations [39] performed in ten dimensional deformation space demonstrate the shapes of a decaying Cf nucleus at large deformations (Fig. 11) in the potential valleys 3 and 4. The distance between the centers of the side constituents ($R_{12}$) are equal to approximately 18 fm and 23 fm.

TABLE III. Neutron – emission probabilities for $^{252}$Cf from [37] and from [38].

| ν | Ref. [37] | Ref. [38] |
|---|---|---|
| 0 | 0.0025±0.0004 | 0.0022±0.0001 |
| 1 | 0.0282±0.0024 | 0.0256±0.0013 |
| 2 | 0.1199±0.0081 | 0.1239±0.0014 |
| 3 | 0.2681±0.0278 | 0.2715±0.0011 |
| 4 | 0.3056±0.0118 | 0.3046±0.0005 |
| 5 | 0.1951±0.0217 | 0.1866±0.0006 |
| 6 | 0.0674±0.0158 | 0.0681±0.0004 |
| 7 | 0.0084±0.0048 | 0.0152±0.0001 |
| 8 | 0.0045±0.0030 | 0.0021±0.0000 |
| 9 | 0.0004±0.0015 | 0.0002±0.0000 |

After the rupture at the narrowest section of the neck, almost all deformation energy (Fig. 11(b)) concentrates in the light (panel c) or heavy fragment (panel d).

Typical shapes of a fissioning nucleus at large deformations are confirmed independently by the neutron data from [40] (Fig. 12).

At high neutron multilicities almost all deformation energy is concentrated in the light (Fig. 12(a)) or the heavy fragment (Fig. 12(b)) in agreement with the shape of the nucleus in the valley 3 and valley 4 respectively. For the $^{248}$Cm, such tendency was traced up to the $v_L/v_H$ = 9/0. Just for a sense of the scale of yields of highly deformed scission configurations, one can cite to the relative total yield $Y_R$ of the fission events at $v_{tot}$ = 6 and $v_L/v_H$ = 6/0. $Y_R$ was estimated to be 2.19 % and 0.72 % for $^{248}$Cm and $^{252}$Cf respectively [41].

We assume that in contrast to conventional ternary fission *the CCT occurs as a two-step decay of an extremely deformed prescission nuclear configuration in the valley of true ternary fission ([6], Fig. 6) or states associated with cold deformed fission in the binary channel. According to the neutron data, the population of such states reaches several percent.*

Presumably, the states of the fissioning nucleus in the valley 3 could give rise to the rectangular structure in Fig. 2(a) consisting of two horizontal lines $M_1$ = 68 and 72 u, and the vertical line $M_2$ = 128 u. At high values of $E^*$, the magic Sn cluster is well preformed in the body of the mother system (Fig. 11(a), valley 3), while the nascent deformed light cluster has enough energy for a brake-up (Fig. 11(b)). Predominantly, the magic clusters of Ni, Ge, Sr, and Mo [2] are formed.

Cluster decay from the excited state [42] might be the mechanism responsible for this "breakup" process.

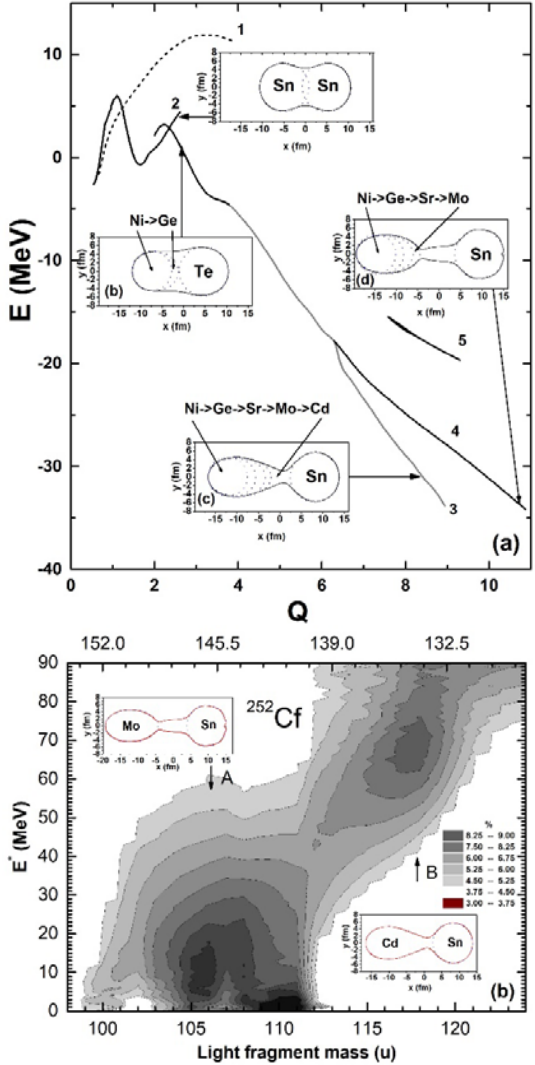

FIG. 10 11. Potential energy of a fissioning $^{252}$Cf nucleus, corresponding to the bottoms of the potential valleys, as a function of $Q$, proportional to its quadrupole moment. The valleys found are marked by numbers 1 to 5. The panels depict the shapes of the system at the points marked by arrows (a). The conditional experimental mass-energy distribution $P(M|E^*)$, where $E^*$ is an excitation energy at the scission point. The panels depict the shapes of the fissioning system following from the calculations ascribed to the two dominant structures [39] (b).

Vertical line $M_2 = 128$ u ($^{128}$Sn) could be the result of such a scenario. At less excitation, the mass of the heavy fragment shifts towards larger masses (Fig. 11(b)), while the light clusters mentioned above, including Ni, continue to form. The lines $M_1 = 68$ and 72 u display this way.

The less populated valley 4 could give rise to the ternary events from the locus $w2$ (Fig. 3(a)). The shape of the fissioning system there suggests that light magic clusters, marked in panel (d), (Fig. 11(a)) become free after the first rupture.

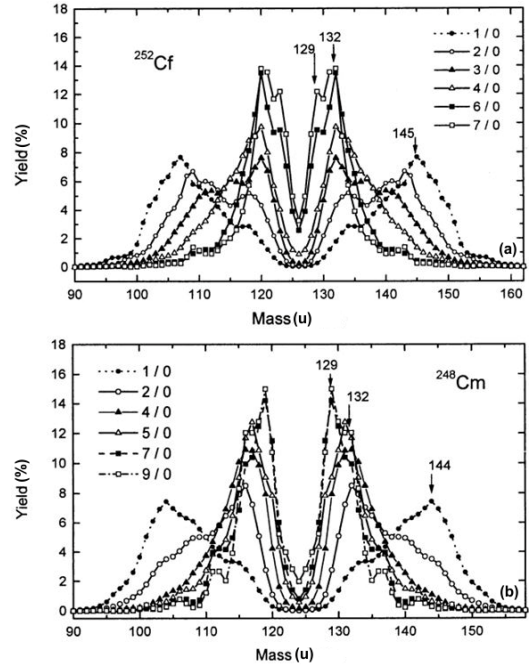

FIG. 12. Partial fragment mass distributions for fixed numbers of emitted neutrons for $^{252}$Cf (a, b), $v_L/v_H$ denotes the number of neutrons emitted from the light and heavy fragments, respectively. The yield is normalized to the total number of events with the neutron multiplicity $v_{tot} = v_L + v_H$ [40].

It appears that there is a profound analogy of the CCT mechanism proposed here and the process of non-equilibrium fission discussed in Ref [43]. Kinematically complete experiments have been performed on two- and three-body exit channel in the reactions $^{84}$Kr + $^{166}$Er and $^{129}$Xe + $^{122}$Sn at 12.5 MeV/u. The FF angular distribution observed was consistent with an orientation of the fission axis, being approximately collinear with the axis of the first scission. A time-scale of $10^{-21}$s between the consecutive scission acts was established. The following scenario of the interaction was proposed: "The primary deep-inelastic collision results *in strongly elongated and mass-asymmetrically deformed fragments* with deformation axes about collinear with the line of separation. *A fraction of up to 10% is so strongly deformed that it directly proceeds toward scission without passing through a shape-equilibrated stage, keeping the memory of the initial orientation*".

**B. Collinearity in the ternary decay.**

The preference of the prolate, chain-like saddle point configurations in fission into three equal fragments was shown first in Refs. [44, 45]. A collinear prescission configuration of the decaying system is only a necessary but not a sufficient condition for collinear tri-partition. There are two factors that could be decisive in non-collinear kinematics of the ternary decay.

### B.1. Ejection of the central fragment from the fission axis by the side fragments.

Obviously, at the scission configuration of the system, shown in Fig. 7(a), at any offset of the $FF_3$ from the fission axis, there is a force, acting on the fragment in the direction perpendicular to the axis. In the notation of Ref. [5], the "true ternary" decay process could be considered only a hypothesis. Only under this hypothesis, the authors came to the expected conclusion that "collinearity is extremely unstable". In contrast, all the scission scenarios agreeing with the experimental results (Table I and fig.9) can be attributed to sequential or almost sequential ones. The dinuclear system formed after the first rupture decays predominantly at $R_{12} > 40$ fm that provides collinear (at the experimental angular resolution) tri-partition [8].

### B.2. Possible rotation of the dinuclear system before its decay.

The origin of the angular momentum of a fission fragment is explained by the excitation of transverse vibrations in a fissioning nucleus before the rupture. A linear increase of the fragments' spins as a function of the fragments excitation energy appears to be closely related to their deformation [46]. According to Figs. 11 and 12, a heavy fragment is born slightly deformed, thus getting low or even zero spin (this is the case, at least for a spherical Sn nucleus). Keeping in mind zero spin of a mother Cf nucleus, one could also expect low spin of the nascent light fragment. However, as it was pointed out in Ref. [47]: "Even for the spontaneous fission of Cf, which has an angular momentum of zero, the products do not have to have identical and canceling angular momentum. Whatever deviations that do exist between the two primary products can be made up by orbital angular momentum of the system." Collective vibrations such as wriggling, tilting, bending, and twisting can bear angular momentum [48], but only if they are really allowed for the specific prescission shape of the nucleus.

Thus, a reliable prediction of the intrinsic angular momentum of the dinuclear system forming after the first rupture seems a nontrivial task. One could expect low and even zero spin; and in any case, the arbitrary extrapolation of the conclusions, typical for binary fission to the CCT, is questionable.

### C. Peculiarities of experimental observation of the CCT.

One of the questions addressed by our experiments [5] is connected to the fact that the Ni-bump is observed predominantly in the spectrometer arm facing the source backing (Fig. 1(a)). Similar effect concerning the yield of ternary fission events is mentioned also in Ref. [49].

The results of our recent experiments on fission fragments, passing through different metal foils [50, 51], allow us to assume that the bulk of the fragments from the conventional binary fission are born in the shape-isomer states. The Coulomb excitation leads to a brake-up of the fragment while it passes the foil. The driving potential for the typical FF [50] shows a general tendency to fusion of the constituents and pronounced minima for the partitions that involve magic clusters. The life-time against fusion is a function of the FF excitation. For the extremely excited FFs in the CCT, this time could be quite short. Even the slight Coulomb excitation, caused by FF scattering at the small angle, occurring just in the source backing, can lead to the FF brake-up.

Thus, the influence of the Coulomb excitation is too small to significantly impact the results of our calculations (Table I) that are consistent with the experiment. It should be stressed that we also have observed the CCT modes showing quite the same manifestation in both spectrometer arms [52].

The next point that needs clarification is the results of the experimental work [53] of A. V. Kravtsov and G. E. Solyakin, dedicated to searching for the collinear tri-partition of $^{252}$Cf. They did not observe any disintegration of the type

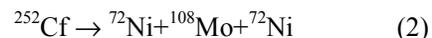
$$^{252}\text{Cf} \rightarrow {}^{72}\text{Ni} + {}^{108}\text{Mo} + {}^{72}\text{Ni} \qquad (2)$$

at the level $7.5 \times 10^{-6}$ for the mass of the central fragment 75 u $< M_3 <$ 152 u under the condition that the side fragments' momenta are similar. The used selection rule excludes the possibility to reveal the events from the Ni-bump (Fig. 3) under discussion.

Even for the configuration (2), in more realistic sequential decay, the momenta of the side fragments would differ radically, and such a mode would be lost as well.

## VI. DISCUSSION

### A. Some facts in support of the CCT.

The "true ternary" decay model, used in Ref. [5], at $x_r = 1$ (Fig. 8(b)) and with the appropriate choice of TXE is similar to the decay scenario which we successfully applied for the analysis of the loci *w*3 (fig.9). Indeed, the predicted energies of all three fragments obtained in Ref. [5] (Fig. 8(b)) are in good agreement with our experimental results for the *w*3 group of events (Figs. 3 and 4: TXE ~ 30 MeV, $E_L$ = 90–110 MeV, $E_T$ < 8 MeV, $E_H$ ~ 80 MeV). It should be mentioned that our experimental results shown in Figs. 3(a) and 3(b) were published [54] three years before the work [5]. Nevertheless, they are not discussed in Ref. [5], and the conclusion of the authors "For true ternary fission, any significant TXE > 0 MeV is therefore not expected" contradicts the experimental data.

Unfortunately, the configuration "Ni centered" was not considered in the "true ternary" decay model. That is why; despite our criticism of Ref. [5], in the implementation of the sequential decay model, its predictions (Fig. 7 in Ref. [5]) in semi-quantitative manner are comparable with our experimental results for the locus *w*1*a* in Fig. 3(c). Experimental values $E_{LF}$ ~20 MeV and $E_{TP}$ ~100 MeV are confirmed.

Returning to the scenario of the CCT process proposed here (Sec. IV), we would like to emphasize that it does not exclude a non-collinear decay of a prescission TNS. Decay of the dinuclear system forming after the first rupture can happen with some probability at $R_{12}$< 40 fm. This value is a conditional border of the inter-cluster distances that provides collinear decay geometry [10]. It is such type of events that possibly were observed in the series of the latest experiments [49, 55].

The mass and energy spectra for $^{252}$Cf ternary fission events obtained in Ref. [49] are shown in Fig. 13. They agree well with those obtained in Ref. [55]. Low-energy threshold of 25 MeV, used in the measurements [49] (compare with Fig. 3(b)), could be a reason of disagreement in the yields of the ternary events in Refs. [49] and [55] respectively. The yield of the lightest peak in Fig. 13(b) is approximately $10^{-6}$ per binary fission. Parameters of the ternary decay products in Fig. 13 correlate well with our data for the loci *w*2 and *w*3 in Fig. 3.

### B. Binary and ternary fission – different fission ways?

Our arguments concerning a possible CCT mechanism along with the experimental results are based on three different theoretical approaches. These approaches are 1) the concept of the trinuclear system [11]; 2) calculation of the potential energy surface (PES) of the fissioning $^{252}$Cf nucleus under the three-center shell model (TSM) [6]; 3) calculation of the PES in ten-dimensional deformation space [39]. Only the last two approaches describe the evolution of the system starting from the ground state. TSM calculations predict a separate valley of ternary fission with additional third hump at the exit (Fig. 6). Last barrier is apparently connected with formation of the second pinch on the neck [56] or preformation of the chain of three partially overlapping nuclei. At the scission point, the masses of all three nascent fragments, which are predominantly magic nuclei, are well defined.

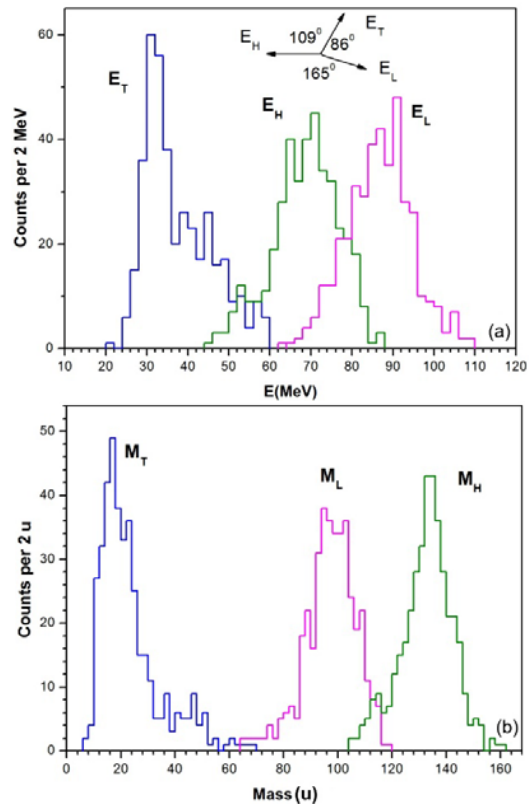

FIG. 13. (Color online). Energy spectrum of $^{252}$Cf ternary fission events, measured with a low-energy threshold of 25 MeV. The inset indicates the corresponding mean angles (a). Mass distribution calculated from the measured energies and angles using momentum conservation [49] (b).

In the alternative approach [39], evolution of a fissioning nucleus leads, at high elongation, to a specific nuclear shape (Fig. 10, valley 3) with a strongly deformed light nascent fragment. After the first rupture, highly excited light fragment could undergo cluster decay. As in the case of cluster radioactivity [34], one of the CCT partners should be a magic nucleus.

The question of which fission way is actually responsible for the CCT remains open. In any case, very deformed prescission configuration is predicted, which agrees with our experimental results and model calculations (Table I, Fig. 9).

## C. Recommendations for independent experimental verification of the CCT.

The most desirable approach to independent experimental verification of the CCT consists in involvement of the methodic alternative to the $2V$–$2E$ method used in our work. In this sense, the mass-separator Lohengrin appears to be a good choice. Moreover, a series of experiments dedicated to observation of the light FFs from the far asymmetric fission, including isotopes of Ni, has been already performed at the Lohengrin (Fig. 14).

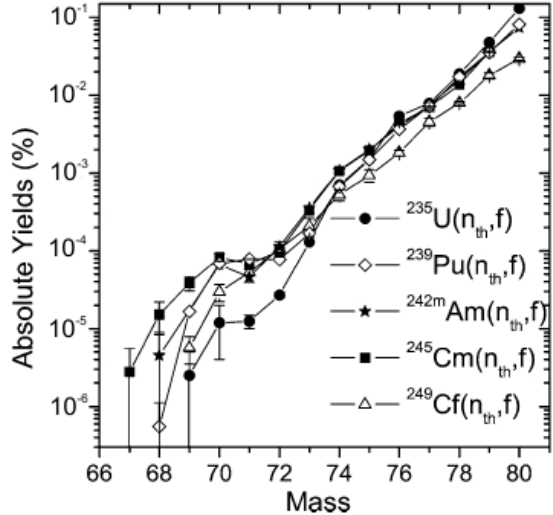

FIG. 14. Absolute fragment-mass yields for the reactions induced by thermal neutrons [57].

For the reaction $^{249}$Cf($n_{th}$, $f$), the yield of the mass 72 u is approximately $10^{-6}$ per binary fission, while the authors of Ref. [5] refer to much lower yield – $10^{-8}$. For the same FF energy range, we have the yield $5\times10^{-5}$ in $^{252}$Cf(sf) (Figs. 3(b) and 3(c)). The shoulder in the mass-yields around the mass number ~70 was also observed in experiments with a twin ionization chamber, and $2E$ method was applied for calculation of the fragment mass of binary fission. The yield of $^{72}$Ni was approximately $10^{-5}$ per binary fission, and its energy was about 104 MeV [26, 27]. The yields of the same order were observed for Ni isotopes in far asymmetric neutron induced fission of $^{236}$U [58].

The energy range of the detected Ni nuclei at the Lohengrin covers energies specific for both binary and ternary fission ($w3a$ group in Fig. 3(c)). Thus, a total yield ~$6\times10^{-5}$ per binary fission could be expected. Strictly speaking, direct comparison of the yields of the fragments close to Ni in spontaneous $^{252}$Cf(sf), and neutron induced fission $^{249}$Cf($n_{th}$, $f$) is not absolutely legitimate. For instance, huge differences in FF mass spectra for $^{258}$Fm(sf) and $^{257}$Fm($n_{th}$, $f$) are well known [59].

In our opinion, registration of the Ni isotopes with the energy lower than 25 MeV, specific for the CCT only (Fig. 3(b)), would be a significant evidence of the existence of this decay channel.

Other candidates for a verification experiment could be the lightest CCT partners in the mass range of 32–50 u (missing masses in the fission events incorporated into the linear structures marked by the arrows 2 and 3 in Fig. 3). Nucleon composition of the corresponding nuclei is determined by the heavier magic partners of ternary decay (see, for instance, ternary partitions shown in bold in Table I). The yield of each composition does not exceed $10^{-6}$ per binary fission. This yield is much higher than that measured in Ref. [30] for some LCP from the mass range under discussion, but such LCP energies are not consistent with those predicted in Fig. 3(c). Searching for the fast LCP seems to be a less promising verification experiment in comparison with the detection of the low energy Ni fragments.

In the Sec. I we discussed multiple problems associated with identification of the CCT events when using time-of-flight spectrometers. The judgment of the authors of Ref. [5] that the effect would have been "easily detected in the past" contradicts to the existence of these objective problems. Two basic characteristics of the effect, namely almost zero energy of one of the decay partners (most often) and almost collinear geometry of their velocities, make detection of all three CCT fragments a complicated methodic problem.

In conclusion, we would recommend registration of the Ni isotopes with the energy lower than 25 MeV at the mass-separator Lohengrin as the primary experiment for verification of the CCT.

## ACKNOWLEDGMENTS

This work was supported, in part, by the Russian Science Foundation and fulfilled in the framework of MEPhI Academic Excellence Project (contract 02.a03.21.0005, 27.08.2013) by the Department of Science and Technology of the Republic of South Africa (RSA) and Bundesministerium für Bildung und Forschung (Germany). We are grateful to N.V. Antonenko, G.G. Adamian, V.I. Furman, A.K. Nasirov, A.V. Karpov, and F.F. Karpeshin for the fruitful discussions.

[1] Yu.V. Pyatkov, D.V. Kamanin, W. von Oertzen, A.A. Alexandrov, I.A. Alexandrova, O.V. Falomkina, N.A. Kondratjev, Yu.N. Kopatch, E.A. Kuznetsova, Yu.E. Lavrova, A.N. Tyukavkin, W. Trzaska, V.E. Zhuchko, Eur. Phys. J. A **45**, 29 (2010).
[2] Yu.V. Pyatkov, D.V. Kamanin, W. von Oertzen, A.A. Alexandrov, I.A. Alexandrova, O.V. Falomkina, N. Jacobs, N.A. Kondratjev, E.A. Kuznetsova, Yu.E. Lavrova, V. Malaza, Yu.V. Ryabov, O.V. Strekalovsky, A.N. Tyukavkin and V.E. Zhuchko, Eur. Phys. J. A **48**, 94 (2012).
[3] D. V. Kamanin, Yu. V. Pyatkov "Clusters in Nuclei – Vol. 3" ed. by C. Beck, Lecture Notes in Physics 875, pp. 183–246 (2013).
[4] H.-G. Ortlepp and FOBOS collaboration, Nucl. Inst. Meth. A **403**, 65 (1998).
[5] P. Holmvall, U. Köster, A. Heinz, and T. Nilsson, Phys. Rev. C **95**, 014602 (2017).
[6] A. V. Karpov, Phys. Rev C **94**, 064615 (2016).
[7] V.I. Zagrebaev, A.V. Karpov, W. Greiner, Phys. Rev. C **81**, 044608 (2010).
[8] V. Zagrebaev and W. Greiner Clusters in Nuclei – Vol. 1" ed. by C. Beck, Lecture Notes in Physics 818, pp. 267–315 (2010).
[9] A. Diaz-Torres, Phys. Lett. B **594**, 69 (2004).
[10] A.K. Nasirov, R.B. Tashkhodjaev and W. von Oertzen, Eur. Phys. J. A **52**, 135 (2016).
[11] R. B. Tashkhodjaev, A. K. Nasirov, and E. K. Alpomeshev, Phys. Rev. C **94**, 054614 (2016).
[12] K.R. Vijayaraghavan and M. Balasubramaniam, W. von Oertzen, Phys. Rev. C **90**, 024601 (2014).
[13] K.R. Vijayaraghavan and M. Balasubramaniam, W. von Oertzen, Phys. Rev. C **91**, 044616 (2015).
[14] W. von Oertzen, A.K. Nasirov, Physics Letters B **734**, 234 (2014).
[15] F. F. Karpeshin, Phys. Atomic Nuclei. **78**(5), 548 (2015).
[16] K.R. Vijiayaraghavan, W. von Oertzen and M. Balasubramaniam, Eur. Phys. J. A **48**, 27 (2012).
[17] Ch. Straede, C. Budtz-Jorgensen, and H.-H. Knitter, Nucl. Phys. A **462**, 85 (1987).
[18] K. Manimaran and M. Balasubramaniam, Eur. Phys. J. A **45**, 293 (2010).
[19] C. Wagemans, "Ternary Fission" in the book "The Nuclear Fission Process" ed. by C. Wagemans, CRC Press.
[20] K. Manimaran and M. Balasubramaniam, Phys. Rev. C **83**, 034609 (2011).
[21] R.B. Tashkhodjaev, A.K. Nasirov, and W. Scheid, Eur. Phys. J. A **47**, 136 (2011).
[22] Yu.V.Pyatkov, D.V.Kamanin, Yu.N.Kopach, A.A.Alexandrov, I.A.Alexandrova, S.B.Borzakov, Yu.N.Voronov, V.E.Zhuchko, E.A.Kuznetsova, Ts.Panteleev, and A. N. Tyukavkin, Physics of Atomic Nuclei, **73**(8), 1309 (2010).
[23] R.B. Tashkhodjaev, A.I. Muminov, A.K. Nasirov, W. von Oertzen, and Yongseok Oh, Phys. Rev. C **91**, 054612 (2015).
[24] V.Yu. Denisov, N.A. Pilipenko, and I.Yu. Sedykh, Phys. Rev. C **95**, 014605 (2017).
[25] S G Rohozinski, Reports on Progress in Physics, **51**(4), 541 (1988).
[26] G. Barreau, A. Sicre, F. Caïtucoli, M. Asghar, T.P. Doan, B. Leroux, G. Martinez and T. Benfoughal, Nucl. Phys. A **432**, 411 (1985).
[27] C. Budtz-Jorgensen and H.-H. Knitter Nucl. Phys. A **490**, 307 (1988).
[28] Tsien San-Tsiang, Ho Zah-Wei, L. Vigneron, and R. Chastel, Phys. Rev. **71**, 382 (1947).
[29] F. Gönnenwein, M. Mutterer, Yu. Kopatch, Europhysics News **36**, 11 (2008).
[30] I. Tsekhanovich, Z. Büyükmumcu, M. Davi, H.O. Denschlag, F. Gönnenwein, and S.F. Boulyga, Phys. Rev. C **67**, 034610 (2003).
[31] A.V. Andreev, G.G. Adamian, N.V. Antonenko, S.P. Ivanova, S.N. Kuklin, and W. Scheid, Eur. Phys. J. A **30**, 579 (2006).
[32] G. Adamian, N. Antonenko, and W. Scheid, "Clusters in Nuclei – Vol. 2" ed. by C. Beck, Lecture Notes in Physics 848, pp. 165–228 (2012).
[33] D.N. Poenaru, W. Greiner, J.H. Hamilton, A.V. Ramayya, E. Hourany, and R.A. Gherghescu, Phys. Rev. C **59**, 3457 (1999).
[34] D. Poenaru and W. Greiner, "Clusters in Nuclei – Vol. 1" ed. by C. Beck, Lecture Notes in Physics 818, pp. 1–56 (2010).
[35] B. Signarbieux, R. Babinet, H. Nifenecker, J. Poitou, Proc. of the 3[th] IAEA Symposium on the Physics and Chemistry of Fission, Rochester, New York, 13–17 August 1973, vol. II, 179-189.
[36] T.D. Thomas, W.M. Gibson, and G.I. Sufford in Symp. Physics and Chemistry of Fission, vol.1, IAEA, Vienna, 1966, 467.
[37] J.F. Wild, J. van Aarle, W. Westmeier, R.W. Lougheed, E.K. Hulet, K.J. Moody, R.J. Dougan, E.-A. Koop, R.E. Glaser, R. Brandt, and P. Patzelt, Phys. Rev. C **41**, 640 (1990).
[38] R. R. Spencer, R. Gwin, and R. Ingle, Nucl. Sci. Eng. **80**, 603 (1982).
[39] Yu.V. Pyatkov, V.V. Pashkevich, Yu.E. Penionzhkevich, V.G. Tishchenko, A.V. Unzhakova, H.-G. Ortlepp, P. Gippner, C.-M. Herbach, W. Wagner, Nucl. Phys. A. **624**, 140 (1997).

[40] V.A. Kalinin, V.N. Dushin, V.A. Jakovlev, B.F. Petrov, A.S. Vorobyev, A.B. Laptev, O.A. Shcherbakov, F.-J. Hambsch, In Proceedings of the "Seminar on Fission Pont D'Oye V", Castle of Pont d' Oye, Habay-la-Neuve, Belgium, 16–19 September, 2003, p. 73–82.

[41] A.S. Vorobiev, Petersburg Institute of nuclear Physics named B.P. Konstantinov, Thesis, 2004 (in Russian, http://gigabaza.ru/doc/2737-all.html).

[42] S.N. Kuklin, G.G. Adamian, and N.V. Antonenko, Physics of Particles and Nuclei, **47**, 206 (2016).

[43] P. Glässel, D.V. Harrach, and H.J. Specht, Z. Phys. A **310**, 189 (1983).

[44] H. Diehl and W. Greiner, Phys. Lett. B **45**, 35 (1973).

[45] H. Diehl and W. Greiner, Nuclear Physics A **229**, 29 (1974).

[46] H. Nifenecker, C. Signarbieux, M. Ribrag, J. Poitou, and J. Matuszek, Nucl. Phys. A **189**, 285 (1972).

[47] J.B. Wilhelmy, E. Cheifetz, R.C. Jared, S.G. Thompson, and H.R. Bowman, Phys. Rev. C **5**, 2041 (1972).

[48] L.G. Moretto and R.P. Schmitt, Phys. Rev. C **21**, 204 (1980).

[49] P. Schall, P. Heeg, M. Mutterer and J.P. Theobald, Phys. Lett. B **191**, 339 (1987).

[50] Yu.V. Pyatkov, D.V. Kamanin, A.A. Alexandrov, I.A. Alexandrova, E.A. Kuznetsova, Yu.E. Lavrova, A.O. Strekalovsky, O.V. Strekalovsky, V.E. Zhuchko, Physics Procedia **74**, 67 (2015).

[51] D.V. Kamanin, Yu.V. Pyatkov, A.A Alexandrov, I.A. Alexandrova, N. Mkaza, V. Malaza, E.A. Kuznetsova, A.O. Strekalovsky, O.V. Strekalovsky and V.E. Zhuchko, Journal of Physics: Conference Series, **863**, 012045 (2017).

[52] Yu.V. Pyatkov, D.V. Kamanin, J.E. Lavrova, N. Mkaza, V. Malaza and A.O. Strekalovsky, Journal of Physics: Conference Series, **863**, 012046 (2017).

[53] A.V. Kravtsov and G.E. Solyakin, Phys. Rev. C, **60**, 017601 (1999).

[54] Yu. V. Pyatkov, D.V. Kamanin, W. von Oertzen, A.A. Alexandrov, I.A. Alexandrova, N.A. Kondratyev, E.A. Kuznetsova, O.V. Strekalovsky, V.E. Zhuchko, in Proceedings of the 20[th] International Seminar on Interaction of Neutrons with Nuclei (ISINN), Dubna, 2013, pp. 104–110.

[55] M.L. Muga, C.R. Rice, and W.A. Sedlacek, Phys. Rev. Lett. **18**, 404 (1967).

[56] V.M. Strutinsky, N.Ya. Lyashchenko and N.A. Popov, Nucl. Phys. **46**, 639 (1963).

[57] D. Rochman, I. Tsekhanovich, F. Gönnenwein, V. Sokolov, F. Storrer, G. Simpson, O. Serot, Nucl. Phys. A **735**, 3 (2004).

[58] A.A. Goverdovsky, V.F. Mitrofanov, V.A. Khrjachkov, Jadernaja Fizika **58**, 1546 (1995) (in Russian).

[59] D.C. Hoffman, J.B. Wilhelmy, J. Weber, W.R. Daniels, E.K. Hulet, R.W. Lougheed, J.H. Landrum, J.F. Wild, and R.J. Dupzyk, Phys. Rev. C **21**, 972 (1980).